\documentclass[a4paper]{PoS}
\usepackage{amsmath}
\usepackage{amssymb}
\usepackage{epsfig}
\usepackage{euscript}
\def\mathswitchr#1{\relax\ifmmode{\mathrm{#1}}\else$\mathrm{#1}$\fi}

\newcommand {\pslash}{\hbox{$\not\hbox{\kern-2.3pt $p$}$}}

\usepackage{cite}

\usepackage{epic}

\def\alf1{ {\alpha\over\pi} }
\def\rQCED{{\rm QCED}}
 
\title{Exact Amplitude-Based Resummation in Quantum Field Theory: Recent Results}
\ShortTitle{Exact Amplitude-Based Resummation in Quantum Field Theory: Recent Results}
\author{\speaker{B.F.L. Ward}%
    \thanks{Work supported in part by D.o.E. grant DE-FG02-09ER41600.}\\
      Baylor University\\
        E-mail:\email{bfl\_ward@baylor.edu}}
\author{S.K. Majhi\\
      Indian Association for the Cultivation of Science\\
        E-mail: \email{tpskm@iacs.res.in}}
\author{S.A. Yost
 \thanks{Work supported in part by 
D.o.E. grant DE-PS02-09ER09-01 and grants from The Citadel Foundation.}\\
      The Citadel\\
        E-mail: \email{scott.yost@citadel.edu}}

\abstract{
We present the current status of the application of our approach of exact amplitude-based resummation in quantum field theory to two areas of investigation: precision QCD calculations of all three of us as needed for LHC physics and the resummed quantum gravity realization by one of us (B.F.L.W.) of Feynman's formulation of Einstein's theory of general relativity. We discuss recent results as they relate to experimental observations. There is reason for optimism in the attendant comparison of theory and experiment.}

\FullConference{10th International Symposium on Radiative Corrections (Applications of Quantum Field Theory to Phenomenology)\\
                 September 26-30, 2011\\
                 Mamallapuram, India\\
                  BU-HEPP-11-04,\\ 
                   Dec., 2011 }

\begin{document}
%

 
\def\Kmax{K_{\rm max}}\def\ieps{{i\epsilon}}\def\rQCD{{\rm QCD}}

\section{\bf Introduction}\label{intro}\par
With the start-up of the LHC the era of precision QCD, 
by which we mean
predictions for QCD processes at the total precision tag of $1\%$ or better,
is upon us and the need for exact, amplitude-based 
resummation of large higher order effects
is paramount. Such resummation allows one to have better than 1\% precision 
as a realistic goal as we shall show in what follows, so that one can indeed 
distinguish new physics(NP) from higher order SM processes and can distinguish 
different models of new physics from one another as well. 
In a parallel development, the issue of the application of ordinary 
quantum field theoretic methods to Einstein's theory of general relativity 
lends itself as well to a resummation approach, provided again that the 
resummation is an exact amplitude-based one, as one of us(B.F.L.W.) 
has shown. In what follows, we present the status of these two 
applications of exact amplitude-based
resummation theory in quantum field theory.\par
The two paradigms which we present are then as follows. First, in the next Section, we present an approach to precision LHC physics 
which is an amplitude-based QED$\otimes$QCD($\equiv\text{QCD}\otimes\text{QED}$) exact resummation 
theory~\cite{qced} 
realized by MC methods. The starting point is then the well-known 
fully differential representation
\begin{equation}
d\sigma =\sum_{i,j}\int dx_1dx_2F_i(x_1)F_j(x_2)d\hat\sigma_{\text{res}}(x_1x_2s)
\label{bscfrla}
\end{equation}
of a hard LHC scattering process
using a standard notation so that the $\{F_j\}$ and 
$d\hat\sigma_{\text{res}}$ are the respective parton densities and 
reduced hard differential cross section where we indicate the that latter 
has been resummed
for all large EW and QCD higher order corrections in a manner consistent
with achieving a total precision tag of 1\% or better for the total 
theoretical precision of (\ref{bscfrla}). The key issue to precision
QCD theory is then the determination of the value of the 
total theoretical precision of (\ref{bscfrla}), which we denote by $\Delta\sigma_{\text{th}}$. It can be decomposed as follows:
\begin{equation}
\Delta\sigma_{\text{th}}= \Delta F \oplus\Delta\hat\sigma_{\text{res}}
\label{eqdecomp1}
\end{equation}
in an obvious notation where $\Delta A$ is the contribution of the uncertainty
on $A$ to $\Delta\sigma_{\text{th}}$.
The theoretical precision $\Delta\sigma_{\text{th}}$ 
validates the application of a given theoretical prediction to precision 
experimental observations, for the discussion of the signals and the
backgrounds for 
both SM and NP studies, and more specifically
for the overall normalization
of the cross sections in such studies. NP can be missed if a calculation
with an unknown value of $\Delta\sigma_{\text{th}}$ 
is used for such studies. This point cannot be emphasized too much.\par
By our definition, $\Delta\sigma_{\text{th}}$ is the 
total theoretical uncertainty coming from the physical precision 
contribution and the technical precision contribution~\cite{jadach-prec}:
the physical precision contribution, $\Delta\sigma^{\text{phys}}_{\text{th}}$,
arises from such sources as missing graphs, approximations to graphs, 
truncations,....; the technical precision contribution, 
$\Delta\sigma^{\text{tech}}_{\text{th}}$, arises from such sources as 
bugs in codes, numerical rounding errors,
convergence issues, etc. The total theoretical error is then 
given by
\begin{equation}
\Delta\sigma_{\text{th}}=\Delta\sigma^{\text{phys}}_{\text{th}}\oplus \Delta\sigma^{\text{tech}}_{\text{th}}.
\end{equation}
The desired value for $\Delta\sigma_{\text{th}}$ depends on the  specific
requirements of the observations. As a general rule, one would 
like that $\Delta\sigma_{\text{th}}\leq f\Delta\sigma_{\text{expt}}$, 
where $\Delta\sigma_{\text{expt}}$ is the respective experimental error
and $f\lesssim \frac{1}{2}$ so that
the theoretical uncertainty does not significantly affect the 
analysis of the data for physics studies in an adverse way.
\par

With the goal of achieving such precision in a provable way, we have 
developed the $\text{QCD}\otimes\text{QED}$ resummation theory in Refs.~\cite{qced}
for the reduced cross section in (\ref{bscfrla}) and for the
resummation of the evolution of the parton densities therein as well.
In both cases, the starting point is the master formula
{\small
\begin{eqnarray}
&d\bar\sigma_{\rm res} = e^{\rm SUM_{IR}(QCED)}
   \sum_{{n,m}=0}^\infty\frac{1}{n!m!}\int\prod_{j_1=1}^n\frac{d^3k_{j_1}}{k_{j_1}} \cr
&\prod_{j_2=1}^m\frac{d^3{k'}_{j_2}}{{k'}_{j_2}}
\int\frac{d^4y}{(2\pi)^4}e^{iy\cdot(p_1+q_1-p_2-q_2-\sum k_{j_1}-\sum {k'}_{j_2})+
D_\rQCED} \cr
&\tilde{\bar\beta}_{n,m}(k_1,\ldots,k_n;k'_1,\ldots,k'_m)\frac{d^3p_2}{p_2^{\,0}}\frac{d^3q_2}{q_2^{\,0}},
\label{subp15b}
\end{eqnarray}}\noindent
where $d\bar\sigma_{\rm res}$ is either the reduced cross section
$d\hat\sigma_{\rm res}$ or the differential rate associated to a
DGLAP-CS~\cite{dglap,cs} kernel involved in the evolution of the $\{F_j\}$ and 
where the {\em new} (YFS-style~\cite{yfs}) {\em non-Abelian} residuals 
$\tilde{\bar\beta}_{n,m}(k_1,\ldots,k_n;k'_1,\ldots,k'_m)$ have $n$ hard gluons and $m$ hard photons and we show the final state with two hard final
partons with momenta $p_2,\; q_2$ specified for a generic $2f$ final state for
definiteness. The infrared functions ${\rm SUM_{IR}(QCED)},\; D_\rQCED\; $
are defined in Refs.~\cite{qced,irdglap1,irdglap2}. This  
simultaneous resummation of QED and QCD large IR effects is exact. Moreover,
the residuals $\tilde{\bar\beta}_{n,m}$ allow a rigorous parton 
shower/ME matching via their shower-subtracted 
counterparts $\hat{\tilde{\bar\beta}}_{n,m}$~\cite{qced}.\par

The result in (\ref{subp15b}) also allows us an an exact, amplitude-based resummation approach to Feynman's formulation of Einstein's theory, as one of us (B.F.L.W.) has  shown in Refs.~\cite{rqg} via
the following representation of the Feynman propagators in that theory: 
\begin{equation}
\begin{split}
i\Delta'_F(k)& =  \frac{i}{(k^2-m^2-\Sigma_s+i\epsilon)}\nonumber\\
&=\frac{ie^{B''_g(k)}}{(k^2-m^2-\Sigma'_s+i\epsilon)}\\
&\equiv i\Delta'_F(k)|_{\text{resummed}}.
\end{split}
\label{rqg1}
\end{equation}
for scalar fields with an attendant generalization for spinning fields~\cite{rqg}. We stress that there are no approximations in (\ref{rqg1}). The formula
for $B''_g(k)$ is given in Refs.~\cite{rqg} and is presented below.
We now discuss the two paradigms opened by (\ref{subp15b}) for precision QCD for the LHC and for exact resummation of Einstein's theory in turn.\par
\section{Precision QCD for the LHC}
We first stress that the methods we emply for resummation of the QCD theory
are fully consistent with the methods in Refs.~\cite{stercattrent1,scet1}. This can be seen by considering the application of the latter methods to the 
$2\rightarrow n$ processes $[f]$ at hard scale Q,
$f_1(p_1,r_1)+f_2(p_2,r_2)\rightarrow f_3(p_3,r_3)+f_4(p_4,r_4)+\cdots+f_{n+2}(p_{n+2},r_{n+2})$, where the $p_i,r_i$ label 4-momenta and color indices respectively, by Abyat {\it et al.} in Ref.~\cite{madg}, where the respective amplitude is represented as
\begin{equation}
\begin{split}
{\cal M}^{[f]}_{\{r_i\}}&=\sum^{C}_{L}{\cal M}^{[f]}_L(c_L)_{\{r_i\}}\\
&= J^{[f]}\sum^{C}_{L}S_{LI}H^{[f]}_I(c_L)_{\{r_i\}},
\end{split}
\label{madg1}
\end{equation}
where repeated indices are summed, and the functions $J^{[f]},S_{LI}$, and $H^{[f]}_I $ are respectively the jet function, the soft function which describes
the exchange of soft gluons between the external lines, and the hard coefficient function. The latter functions' infrared and collinear 
poles have been calculated to 2-loop order in Refs.~\cite{madg}. 
To make contact between eqs.(\ref{subp15b},\ref{madg1}), identify in the
specific application
$\bar{Q}'Q\rightarrow \bar{Q}'''Q''+m(G)$ in (\ref{subp15b}) $f_1=Q, f_2=\bar{Q}',
f_3=Q'', f_4=\bar{Q}''', \{f_5,\cdots,f_{n+2}\}=\{G_1,\cdots,G_m\}$, in (\ref{madg1}), where we use the obvious notation for the gluons here. This means that
$n=m+2$. Then, to use eq.(\ref{madg1}) in eq.(\ref{subp15b}), 
one observes the following:\begin{description}
\item{I.} By its definition in eq.(2.23) of Ref.~\cite{madg}, the anomalous dimension
of the matrix $S_{LI}$ does not contain any of the diagonal effects described by our infrared functions ${\rm SUM_{IR}(QCD)}$ and $D_{\rm QCD}$, where
\[ {\rm SUM_{IR}(QCD)}=2\alpha_s \Re B_{QCD}+2\alpha_s\tilde B_{QCD}(\Kmax),\]
\[ 2\alpha_s\tilde B_{QCD}(\Kmax)=\int{d^3k\over k^0}\tilde S_\rQCD(k)
\theta(\Kmax-k),\]
 \begin{equation} D_\rQCD=\int{d^3k\over k}\tilde S_\rQCD(k)
\left[e^{-iy\cdot k}-\theta(\Kmax-k)\right],\label{subp11a}\end{equation}
where the real IR emission function 
$\tilde S_{\rm QCD}(k)$ and the virtual IR function $\Re B_{QCD}$
are defined eqs.(77,73) in Ref.~\cite{irdglap1}. Note that (\ref{subp15b})
is independent of $K_{max}$.
\item{II.} By its definition in eqs.(2.5) and (2.7) of Ref.~\cite{madg}, the jet function $J^{[f]}$ contains the exponential of the virtual infrared function $\alpha_s\Re{B}_{QCD}$, so that we have to take care that we do not double count when we
use (\ref{madg1}) in (\ref{subp15b}) and in the equations in 
Refs.~\cite{qced,irdglap1,irdglap2} that lead thereto.\end{description}
In this way we get the following realization
of our approach using the results in Ref.~\cite{madg}:
In our result in eq.(75) in Ref.~\cite{irdglap1} for the contribution
to (\ref{subp15b}) of $m$-hard gluons for the process under study here, 
\begin{eqnarray}
  d\hat\sigma^m = {e^{2\alpha_s\Re B_{QCD}}\over {m !}}\int\prod_{j=1}^m
{d^3k_j\over (k_j^2+\lambda^2)^{1/2}}\delta(p_1+q_1-p_2-q_2-\sum_{i=1}^mk_i)
\nonumber\\       
\bar\rho^{(m)}(p_1,q_1,p_2,q_2,k_1,\cdots,k_m)
{d^3p_2d^3q_2\over p^0_2 q^0_2},
\label{diff1}
\end{eqnarray}
we can identify the residual $\bar\rho^{(m)}$ as follows:{\small
\begin{equation}
\begin{split}\bar\rho^{(m)}(p_1,q_1,p_2,q_2,k_1,\cdots,k_m)
&=\overline\sum_{colors,spin}|{\cal M}^{[f]}_{\{r_i\}}|^2\\
&\kern-2cm\equiv \sum_{spins,\{r_i\},\{r'_i\}}\mathfrak{h}^{cs}_{\{r_i\}\{r'_i\}}|\bar{J}^{[f]}|^2\sum^{C}_{L=1}\sum^{C}_{L'=1}S^{[f]}_{LI}H^{[f]}_I(c_L)_{\{r_i\}}\left(S^{[f]}_{L'I'}H^{[f]}_{I'}(c_{L'})_{\{r'_i\}}\right)^\dagger,
\end{split}
\label{madg2}
\end{equation}}
where here we defined $\bar{J}^{[f]}=e^{-\alpha_s\Re{B}_{QCD}}J^{[f]}$, 
and we introduced the color-spin density matrix for the initial state, $\mathfrak{h}^{cs}$, so that
$\mathfrak{h}^{cs}_{\{r_i\}\{r'_i\}}=\mathfrak{h}^{cs}_{\{r_1,r_2\}\{r'_1,r'_2\}}$, suppressing the spin indices, i.e., $\mathfrak{h}^{cs}$ only depends on the initial state colors and has the obvious normalization implied by (\ref{diff1}). Proceeding then according to
the steps in Ref.~\cite{irdglap1} leading from (\ref{diff1}) to 
(\ref{subp15b}) restricted to QCD, we get the corresponding implementation of the results in Ref.~\cite{madg} in our approach, without
any double counting of effects. This proves that the new non-Abelian
residuals $\tilde{\bar\beta}_{m,n}$ in (\ref{subp15b}) transcend those
of an Abelian massless gauge theory as introduced in Ref.~\cite{yfs}.\par 
As we have explained in Refs.~\cite{qced}, these new non-Abelian residuals 
allow rigorous shower/ME matching via their shower subtracted analogs:
\begin{equation}
\tilde{\bar\beta}_{m,n}\rightarrow \hat{\tilde{\bar\beta}}_{m,n}
\end{equation}
where the $\hat{\tilde{\bar\beta}}_{m,n}$ have had all effects in the showers
associated to the $\{F_j\}$ removed from them.
\par 
When the formula in (\ref{subp15b}) is applied to the
calculation of the kernels, $P_{AB}$, in the DGLAP-CS theory itself, 
we get an improvement
of the IR limit of these kernels, an IR-improved DGLAP-CS theory~\cite{irdglap1,irdglap2} in which large IR effects are resummed for the kernels themselves.
The resulting new resummed kernels, $P^{exp}_{AB}$ as given in Ref.~\cite{irdglap1,irdglap2} and as illustrated below, yield a new resummed scheme for the PDF's and the reduced cross section: 
\begin{equation}
\begin{split}
F_j,\; \hat\sigma &\rightarrow F'_j,\; \hat\sigma'\; \text{for}\nonumber\\
P_{gq}(z)&\rightarrow P^{\text{exp}}_{gq}(z)=C_FF_{YFS}(\gamma_q)e^{\frac{1}{2}\delta_q}\frac{1+(1-z)^2}{z}z^{\gamma_q}, \text{etc.},
\end{split}
\end{equation}
with the same value for $\sigma$ in (\ref{bscfrla}) with improved MC stability
as discussed in Ref.~\cite{herwiri}. Here, the YFS~\cite{yfs} infrared factor 
is given by $F_{YFS}(a)=e^{-C_Ea}/\Gamma(1+a)$ where $C_E$ is Euler's constant
and we refer the reader to Ref.~\cite{irdglap1,irdglap2} for the definition of the infrared exponents $\gamma_q,\; \delta_q$ as well as for the complete
set of equations for the new $P^{exp}_{AB}$. $C_F$ is the quadratic Casimir invariant for the quark color representation.\par
The basic physical idea underlying the new kernels is illustrated in Fig.~\ref{fig-bn-1} as it was already shown by Bloch and Nordsieck~\cite{bn1}: 
\begin{figure}[h]
\begin{center}
\epsfig{file=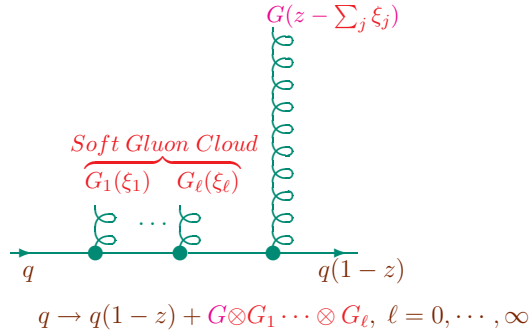,width=70mm}
\end{center}
\label{fig-bn-1}
\caption{Bloch-Nordsieck soft quanta for an accelerated charge.}
\end{figure}
an accelerated charge generates a coherent state of very soft massless quanta of the respective gauge field so that one cannot know which of the infinity of possible states
one has made in the splitting process $q(1)\rightarrow q(1-z)+G\otimes G_1\cdots\otimes G_\ell,\; \ell=0,\cdots,\infty$ illustrated in Fig.~\ref{fig-bn-1}.
The new kernels take this effect into account.\par
\begin{figure}[h]
\begin{center}
\setlength{\unitlength}{1cm}
\begin{picture}(10,5.5)(0,0)
\put(-2,0){\includegraphics[width=60mm]{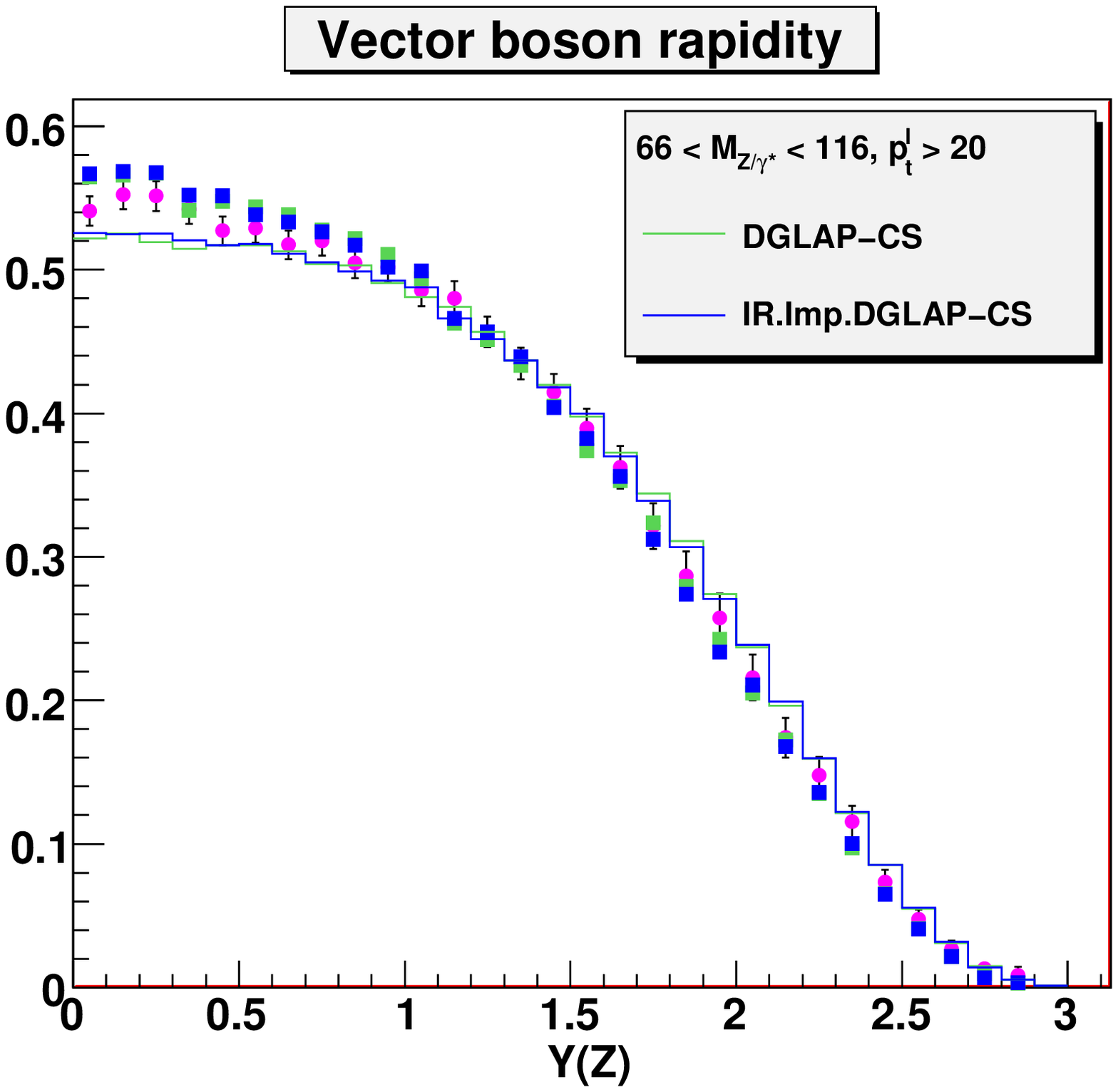}}
\put(5,0){\includegraphics[width=60mm]{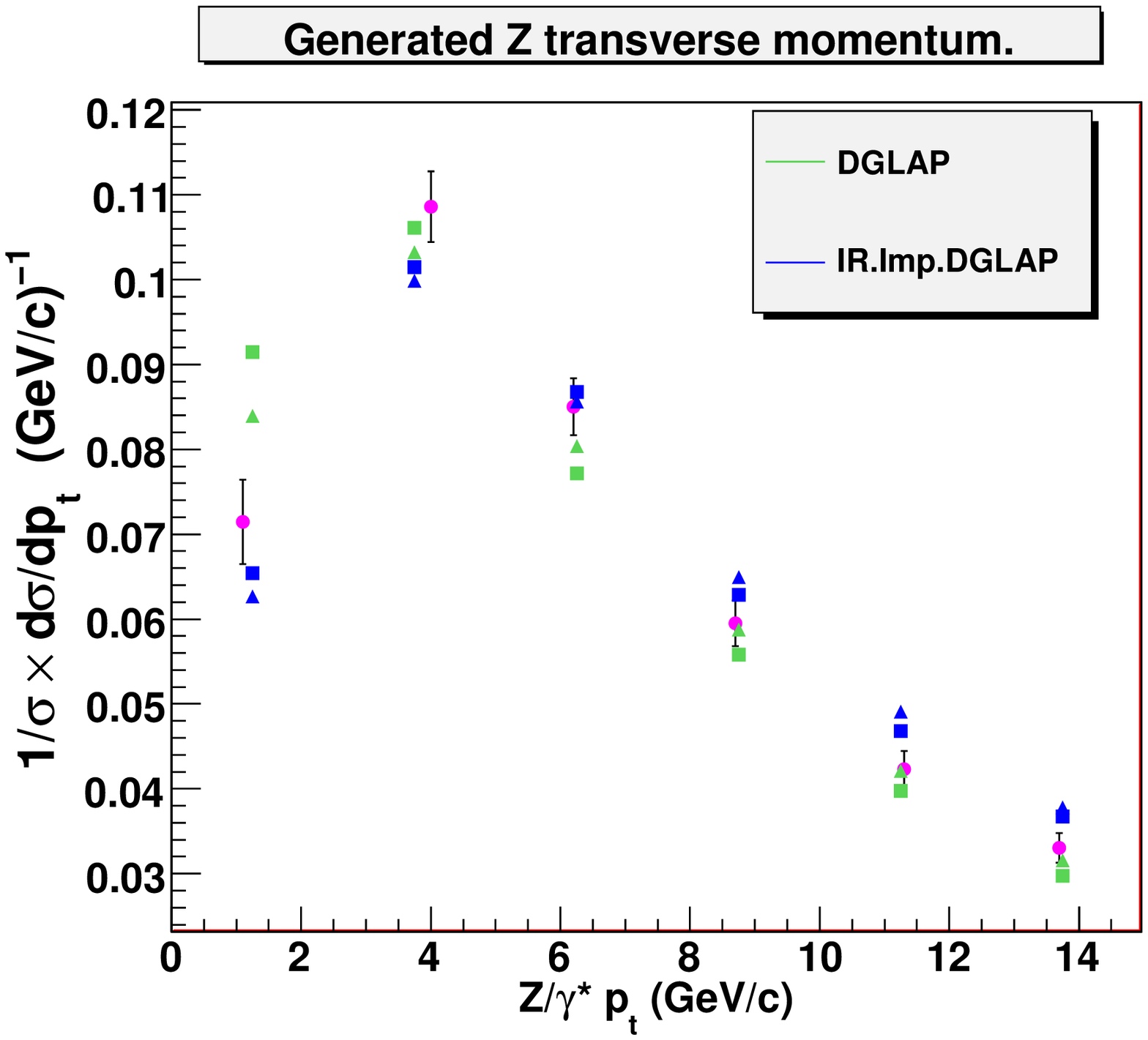}}
\put(1,5.4){\small{(a)}}
\put(7.5,5.4){\small{(b)}}
\end{picture}
\end{center} 
\caption{\baselineskip=11pt  From Ref.~\cite{herwiri}, comparison with FNAL data: (a), CDF rapidity data on
($Z/\gamma^*$) production to $e^+e^-$ pairs, the circular dots are the data, 
the light(dark) 
lines are HERWIG6.510(HERWIRI1.031); 
(b), D0 $p_T$ spectrum data on ($Z/\gamma^*$) production to $e^+e^-$ pairs,
the circular dots are the data, the dark 
triangles are HERWIRI1.031, the light 
triangles are HERWIG6.510. In both (a) and (b) the dark 
squares are MC@NLO/HERWIRI1.031, and the light 
squares are MC@NLO/HERWIG6.510, where MC@NLO/X denotes the realization by MC@NLO of the exact ${\cal O}(\alpha_s)$ correction for the generator X. These are untuned theoretical results.}
\label{fig-hwri1}
\end{figure}

The new MC Herwiri1.031~\cite{herwiri} gives the first realization of the new IR-improved kernels in the Herwig6.5~\cite{hrwg} environment. Here, 
we compare it with Herwig6.510, both with and without
the MC@NLO~\cite{mcnlo} exact ${\cal O}(\alpha_s)$ correction, in Fig.~\ref{fig-hwri1} in relation
to D0 data~\cite{d0pt} on the Z boson $p_T$ in single Z production 
and the CDF data~\cite{galea} on the Z boson rapidity in the same process all at 
the Tevatron. We see~\cite{herwiri} 
that the IR improvement improves the $\chi^2/d.o.f$ 
in comparison with the data in both cases for the soft $p_T$ data and that
for the rapidity data it improves the $\chi^2/d.o.f$ before the application
of the MC@NLO exact ${\cal O}(\alpha_s)$ correction 
and that with the latter correction
the $\chi^2/d.o.f$'s are statistically indistinguishable. More importantly, 
this theoretical paradigm can be systematically improved in principle to reach any desired $\Delta\sigma_{\text{th}}$. The suggested accuracy at the 10\% level shows
the need for the NNLO extension of MC@NLO, in view of our goals
for this process. We are currently developing the analogous
applications for the new kernels for Herwig++~\cite{herpp}, Herwiri++,
for Pyhtia8~\cite{pyth8} and for Sherpa~\cite{shrpa}. In addition
we are currently analysizing recent LHC data using Herwiri1.031/MC@NLO,
Herwiri++/Powheg~\cite{pwhg}
as we shall report elsewhere~\cite{elswh}.\par
\section{Resummed Quantum Gravity}
One of us(B.F.L.W.) has recently continued his application of exact amplitude-based resummation theory to Feynman's formulation of Einstein's theory, as described in Refs.~\cite{rqg}. In particular, in Ref.~\cite{bw-lambda}, he has arrived at a first principles
prediction of the cosmological constant that is close to the observed value~\cite{cosm1,pdg2008}, $\rho_\Lambda\cong (2.368\times 10^{-3}eV(1\pm 0.023))^4$, 
as we now recapitulate.\par
In Ref.~\cite{bw-lambda}, using the deep UV result
\begin{equation} 
B''_g(k)=\frac{\kappa^2|k^2|}{8\pi^2}\ln\left(\frac{m^2}{m^2+|k^2|}\right),       
\label{yfs1} 
\end{equation}
it is shown that the UV limit of Newton's constant, $G_N(k)$, is given by
\begin{equation}
g_*=\lim_{k^2\rightarrow \infty}k^2G_N(k^2)=\frac{360\pi}{c_{2,eff}}\cong 0.0442,
\end{equation} 
where~\cite{rqg,bw-lambda} $c_{2,eff}\cong 2.56\times 10^4$ for the known world.
In addition, it is shown that the contribution of a scalar field to $\Lambda$
is
\begin{equation}
\begin{split}
\Lambda_s&=-8\pi G_N\frac{\int d^4k}{2(2\pi)^4}\frac{(2k_0^2)e^{-\lambda_c(k^2/(2m^2))\ln(k^2/m^2+1)}}{k^2+m^2}\cr
&\cong -8\pi G_N\left[\frac{1}{G_N^{2}64\rho^2}\right],\cr
\label{lambscalar}
\end{split}
\end{equation} 
where $\rho=\ln\frac{2}{\lambda_c}$ 
and we have used the calculus
of Refs.~\cite{rqg,bw-lambda}. We note that 
the standard equal-time (anti-)commutation 
relations algebra realizations
then show that a Dirac fermion contributes $-4$ times $\Lambda_s$ to
$\Lambda$. 
The deep UV limit of $\Lambda$ then becomes
\begin{equation}
\begin{split}
\Lambda(k) &\operatornamewithlimits{\longrightarrow}_{k^2\rightarrow \infty} k^2\lambda_*,\cr
\lambda_*&=-\frac{c_{2,eff}}{2880}\sum_{j}(-1)^{F_j}n_j/\rho_j^2\cr
&\cong 0.0817
\end{split}
\end{equation} 
where $F_j$ is the fermion number of $j$ and $\rho_j=\rho(\lambda_c(m_j))$.
Our results for $(g_*,\lambda_*)$ agree qualitatively with those in Refs.~\cite{reutera,reuter1}.
\par
For reference, we note that, if we restrict our resummed quantum gravity
calculations above for $g_*,\lambda_*$ to 
the pure gravity theory with no SM matter fields, we get the results
$$g_*=.0533,\;\lambda_*=-.000189.$$ We see that our results suggest
that there is still significant cut-off effects in the results 
used for $g_*,\;\lambda_*$\footnote{In the first paper in Ref.~\cite{reuter1}, 
$(g_*,\lambda_*)\approx (0.27,0.36)$.} in Refs.~\cite{reutera,reuter1}, which already
seem to include an effective matter contribution when viewed from
our resummed quantum gravity perspective, as an artifact of the obvious gauge and cut-off dependencies of the results. Indeed, 
from a purely quantum field theoretic point of view,
the cut-off action is 
\begin{equation}
\Delta_kS(h,C,\bar{C};\bar{g})=\frac{1}{2}<h,{\cal R}^{\text{grav}}_kh>+<\bar{C},{\cal R}^{\text{gh}}_kC>
\end{equation}
where $\bar{g}$ is the general background metric, which is the Minkowski space metric
$\eta$ here, 
and $C,\bar{C}$ are the
ghost fields and the operators ${\cal R}^{\text{grav}}_k,\; {\cal R}^{\text{gh}}_k$ implement the course graining as they satisfy the limits 
\begin{equation}
\begin{split}
{\underset{p^2/k^2\rightarrow \infty}{\text{lim}}} {\cal R}_k &=0,\nonumber\\
{\underset{p^2/k^2\rightarrow 0}{\text{lim}}}{\cal R}_k&\rightarrow \mathfrak{Z}_k k^2,
\end{split}
\end{equation}
for some $\mathfrak{Z}_k$~\cite{reutera}. Here, the inner product is that defined
in the second paper in Refs.~\cite{reutera} in its Eqs.(2.14,2.15,2.19).
The result is that the modes with $p\lesssim k$ have a shift of their vacuum energy
by the cut-off operator. There is no disagreement in principle between
our gauge invariant, cut-off independent results and the gauge dependent, cut-off dependent results in Refs.~\cite{reutera,reuter1}.\par
\subsection{An Estimate of $\Lambda$}
To estimate the value of $\Lambda$ today, we take the normal-ordered form of Einstein's equation, 
\begin{equation}
:G_{\mu\nu}:+\Lambda :g_{\mu\nu}:=-8\pi G_N :T_{\mu\nu}:.
\label{eineq2}
\end{equation}
The coherent state representation of the thermal density matrix then gives
the Einstein equation in the form of thermally averaged quantities with
$\Lambda$ given by our result above in lowest order. 
Taking the transition time between the Planck regime and the classical Friedmann-Robertson-Walker regime at $t_{tr}\sim 25 t_{Pl}$ from Refs.~\cite{reuter1},
we introduce
\begin{equation}
\begin{split}
\rho_\Lambda(t_{tr})&\equiv\frac{\Lambda(t_{tr})}{8\pi G_N(t_{tr})}\cr
         &=\frac{-M_{Pl}^4(k_{tr})}{64}\sum_j\frac{(-1)^Fn_j}{\rho_j^2}
\end{split}
\end{equation}
and use the arguments in Refs.~\cite{branch-zap} ($t_{eq}$ is the time of matter-radiation equality) to get 
\begin{equation}
\begin{split}
\rho_\Lambda(t_0)&\cong \frac{-M_{Pl}^4(1+c_{2,eff}k_{tr}^2/(360\pi M_{Pl}^2))^2}{64}\sum_j\frac{(-1)^Fn_j}{\rho_j^2}\cr
          &\qquad\quad \times \big[\frac{t_{tr}^2}{t_{eq}^2} \times (\frac{t_{eq}^{2/3}}{t_0^{2/3}})^3\big]\cr
          &\cong \frac{-M_{Pl}^2(1.0362)^2(-9.197\times 10^{-3})}{64}\frac{(25)^2}{t_0^2}\cr
   &\cong (2.400\times 10^{-3}eV)^4.
\end{split}
\end{equation}
where we take the age of the universe to be $t_0\cong 13.7\times 10^9$ yrs. 
In the latter estimate, the first factor in the square bracket comes from the period from
$t_{tr}$ to $t_{eq}$ (radiation dominated) and the second factor
comes from the period from $t_{eq}$ to $t_0$ (matter dominated)
\footnote{The method of the operator field forces the vacuum energies to follow the same scaling as the non-vacuum excitations.}.
This estimate should be compared with the experimental result~\cite{cosm1,pdg2008}\footnote{See also Ref.~\cite{sola2} for an analysis that suggests a value for $\rho_\Lambda(t_0)$ that is qualitatively similar to this experimental result.}
$\rho_\Lambda(t_0)|_{\text{expt}}\cong (2.368\times 10^{-3}eV(1\pm 0.023))^4$.
In closing, two of us (B.F.L.W., S.A.Y.)
thank Prof. Ignatios Antoniadis for the support and kind 
hospitality of the CERN TH Unit while part of this work was completed.
\par

\end{document}